
\font\twelverm=cmr10 scaled 1200    \font\twelvei=cmmi10 scaled 1200
\font\twelvesy=cmsy10 scaled 1200   \font\twelveex=cmex10 scaled 1200
\font\twelvebf=cmbx10 scaled 1200   \font\twelvesl=cmsl10 scaled 1200
\font\twelvett=cmtt10 scaled 1200   \font\twelveit=cmti10 scaled 1200
\skewchar\twelvei='177   \skewchar\twelvesy='60
\def\twelvepoint{\normalbaselineskip=12.4pt
  \abovedisplayskip 12.4pt plus 3pt minus 9pt
  \belowdisplayskip 12.4pt plus 3pt minus 9pt
  \abovedisplayshortskip 0pt plus 3pt
  \belowdisplayshortskip 7.2pt plus 3pt minus 4pt
  \smallskipamount=3.6pt plus1.2pt minus1.2pt
  \medskipamount=7.2pt plus2.4pt minus2.4pt
  \bigskipamount=14.4pt plus4.8pt minus4.8pt
  \def\rm{\fam0\twelverm}          \def\it{\fam\itfam\twelveit}%
  \def\sl{\fam\slfam\twelvesl}     \def\bf{\fam\bffam\twelvebf}%
  \def\mit{\fam 1}                 \def\cal{\fam 2}%
  \def\tt{\twelvett}
  \textfont0=\twelverm   \scriptfont0=\tenrm   \scriptscriptfont0=\sevenrm
  \textfont1=\twelvei    \scriptfont1=\teni    \scriptscriptfont1=\seveni
  \textfont2=\twelvesy   \scriptfont2=\tensy   \scriptscriptfont2=\sevensy
  \textfont3=\twelveex   \scriptfont3=\twelveex  \scriptscriptfont3=\twelveex
  \textfont\itfam=\twelveit
  \textfont\slfam=\twelvesl
  \textfont\bffam=\twelvebf \scriptfont\bffam=\tenbf
  \scriptscriptfont\bffam=\sevenbf
  \normalbaselines\rm}

\def\beginlinemode{\endmode
  \begingroup\parskip=0pt \obeylines\def\\{\par}\def\endmode{\par\endgroup}}
\def\beginparmode{\endmode
  \begingroup \def\endmode{\par\endgroup}}
\let\endmode=\par
{\obeylines\gdef\
{}}
\def\singlespace{\baselineskip=\normalbaselineskip}
\def\oneandahalfspace{\baselineskip=\normalbaselineskip
  \multiply\baselineskip by 3 \divide\baselineskip by 2}
\def\doublespace{\baselineskip=\normalbaselineskip \multiply\baselineskip by 2}
\newcount\firstpageno
\firstpageno=2
\footline={\ifnum\pageno<\firstpageno{\hfil}\else{\hfil\twelverm\folio\hfil}\fi}
\let\rawfootnote=\footnote              
\def\footnote#1#2{{\rm\singlespace\parindent=0pt\rawfootnote{#1}{#2}}}
\def\raggedcenter{\leftskip=2em plus 12em \rightskip=\leftskip
  \parindent=0pt \parfillskip=0pt \spaceskip=.3333em \xspaceskip=.5em
  \pretolerance=9999 \tolerance=9999
  \hyphenpenalty=9999 \exhyphenpenalty=9999 }
\parskip=\medskipamount
\twelvepoint            
\overfullrule=0pt       
\def\preprintno#1{
 \rightline{\rm #1}}    
\def\author                     
  {\vskip 3pt plus 0.2fill \beginlinemode
   \singlespace \raggedcenter \twelvesc}
\def\affil                      
  {\vskip 3pt plus 0.1fill \beginlinemode
   \oneandahalfspace \raggedcenter \sl}
\def\abstract                   
  {\vskip 3pt plus 0.3fill \beginparmode
   \doublespace \narrower \noindent ABSTRACT: }
\def\endtitlepage               
  {\endpage                     
   \body}
\def\body                       
  {\beginparmode}               

\def\subhead#1{                 
  \vskip 0.1truein             
  {\raggedcenter #1 \par}
   \nobreak\vskip 0.1truein\nobreak}
\def\refto#1{$|{#1}$}           
\def\references                 
  {\subhead{References}         
   \beginparmode
   \frenchspacing \parindent=0pt \leftskip=1truecm
   \parskip=8pt plus 3pt \everypar{\hangindent=\parindent}}
\gdef\refis#1{\indent\hbox to 0pt{\hss#1.~}}    
\gdef\journal#1, #2, #3, 1#4#5#6{               
    {\sl #1~}{\bf #2}, #3, (1#4#5#6)}           
\def\refstylenp{                
  \gdef\refto##1{ [##1]}                                
  \gdef\refis##1{\indent\hbox to 0pt{\hss##1)~}}        
  \gdef\journal##1, ##2, ##3, ##4 {                     
     {\sl ##1~}{\bf ##2~}(##3) ##4 }}
\def\refstyleprnp{              
  \gdef\refto##1{ [##1]}                                
  \gdef\refis##1{\indent\hbox to 0pt{\hss##1)~}}        
  \gdef\journal##1, ##2, ##3, 1##4##5##6{               
    {\sl ##1~}{\bf ##2~}(1##4##5##6) ##3}}

\def\prpts{\journal Phys. Rep., }

\def\endreferences{\body}
\def\endpage                    
  {\vfill\eject}
\def\endpaper                   
  {\endmode\vfill\supereject}
\def\endit
  {\endpaper\end}
\def\ref#1{Ref. #1}                     
\def\Ref#1{Ref. #1}                     

\def\m@th{\mathsurround=0pt }
\font\twelvesc=cmcsc10 scaled 1200
\def\cite#1{{#1}}
\def\(#1){(\call{#1})}
\def\call#1{{#1}}
\def\taghead#1{}
\def\leaderfill{\leaders\hbox to 1em{\hss.\hss}\hfill}
\def\square{\kern1pt\vbox{\hrule height 1.2pt\hbox{\vrule width 1.2pt\hskip 3pt
   \vbox{\vskip 6pt}\hskip 3pt\vrule width 0.6pt}\hrule height 0.6pt}\kern1pt}
\catcode`@=11
\newcount\tagnumber\tagnumber=0

\immediate\newwrite\eqnfile
\newif\if@qnfile\@qnfilefalse
\def\write@qn#1{}
\def\writenew@qn#1{}
\def\w@rnwrite#1{\write@qn{#1}\message{#1}}
\def\@rrwrite#1{\write@qn{#1}\errmessage{#1}}

\def\taghead#1{\gdef\t@ghead{#1}\global\tagnumber=0}
\def\t@ghead{}

\expandafter\def\csname @qnnum-3\endcsname
  {{\t@ghead\advance\tagnumber by -3\relax\number\tagnumber}}
\expandafter\def\csname @qnnum-2\endcsname
  {{\t@ghead\advance\tagnumber by -2\relax\number\tagnumber}}
\expandafter\def\csname @qnnum-1\endcsname
  {{\t@ghead\advance\tagnumber by -1\relax\number\tagnumber}}
\expandafter\def\csname @qnnum0\endcsname
  {\t@ghead\number\tagnumber}
\expandafter\def\csname @qnnum+1\endcsname
  {{\t@ghead\advance\tagnumber by 1\relax\number\tagnumber}}
\expandafter\def\csname @qnnum+2\endcsname
  {{\t@ghead\advance\tagnumber by 2\relax\number\tagnumber}}
\expandafter\def\csname @qnnum+3\endcsname
  {{\t@ghead\advance\tagnumber by 3\relax\number\tagnumber}}

\def\equationfile{%
  \@qnfiletrue\immediate\openout\eqnfile=\jobname.eqn%
  \def\write@qn##1{\if@qnfile\immediate\write\eqnfile{##1}\fi}
  \def\writenew@qn##1{\if@qnfile\immediate\write\eqnfile
    {\noexpand\tag{##1} = (\t@ghead\number\tagnumber)}\fi}
}

\def\callall#1{\xdef#1##1{#1{\noexpand\call{##1}}}}
\def\call#1{\each@rg\callr@nge{#1}}

\def\each@rg#1#2{{\let\thecsname=#1\expandafter\first@rg#2,\end,}}
\def\first@rg#1,{\thecsname{#1}\apply@rg}
\def\apply@rg#1,{\ifx\end#1\let\next=\relax%
\else,\thecsname{#1}\let\next=\apply@rg\fi\next}

\def\callr@nge#1{\calldor@nge#1-\end-}
\def\callr@ngeat#1\end-{#1}
\def\calldor@nge#1-#2-{\ifx\end#2\@qneatspace#1 %
  \else\calll@@p{#1}{#2}\callr@ngeat\fi}
\def\calll@@p#1#2{\ifnum#1>#2{\@rrwrite{Equation range #1-#2\space is bad.}
\errhelp{If you call a series of equations by the notation M-N, then M and
N must be integers, and N must be greater than or equal to M.}}\else%
 {\count0=#1\count1=#2\advance\count1
by1\relax\expandafter\@qncall\the\count0,%
  \loop\advance\count0 by1\relax%
    \ifnum\count0<\count1,\expandafter\@qncall\the\count0,%
  \repeat}\fi}

\def\@qneatspace#1#2 {\@qncall#1#2,}
\def\@qncall#1,{\ifunc@lled{#1}{\def\next{#1}\ifx\next\empty\else
  \w@rnwrite{Equation number \noexpand\(>>#1<<) has not been defined yet.}
  >>#1<<\fi}\else\csname @qnnum#1\endcsname\fi}

\let\eqnono=\eqno
\def\eqno(#1){\tag#1}
\def\tag#1$${\eqnono(\displayt@g#1 )$$}

\def\aligntag#1\endaligntag
  $${\gdef\tag##1\\{&(##1 )\cr}\eqalignno{#1\\}$$
  \gdef\tag##1$${\eqnono(\displayt@g##1 )$$}}

\def\eqalignno#1{\displ@y \tabskip\centering
  \halign to\displaywidth{\hfil$\displaystyle{##}$\tabskip\z@skip
    &$\displaystyle{{}##}$\hfil\tabskip\centering
    &\llap{$\displayt@gpar##$}\tabskip\z@skip\crcr
    #1\crcr}}

\def\displayt@gpar(#1){(\displayt@g#1 )}

\def\displayt@g#1 {\rm\ifunc@lled{#1}\global\advance\tagnumber by1
        {\def\next{#1}\ifx\next\empty\else\expandafter
        \xdef\csname @qnnum#1\endcsname{\t@ghead\number\tagnumber}\fi}%
  \writenew@qn{#1}\t@ghead\number\tagnumber\else
        {\edef\next{\t@ghead\number\tagnumber}%
        \expandafter\ifx\csname @qnnum#1\endcsname\next\else
        \w@rnwrite{Equation \noexpand\tag{#1} is a duplicate number.}\fi}%
  \csname @qnnum#1\endcsname\fi}

\def\ifunc@lled#1{\expandafter\ifx\csname @qnnum#1\endcsname\relax}

\let\@qnend=\end\gdef\end{\if@qnfile
\immediate\write16{Equation numbers written on []\jobname.EQN.}\fi\@qnend}

\catcode`@=12
\refstyleprnp
\catcode`@=11
\newcount\r@fcount \r@fcount=0
\def\refreset{\global\r@fcount=0}
\newcount\r@fcurr
\immediate\newwrite\reffile
\newif\ifr@ffile\r@ffilefalse
\def\w@rnwrite#1{\ifr@ffile\immediate\write\reffile{#1}\fi\message{#1}}

\def\writer@f#1>>{}
\def\referencefile{
  \r@ffiletrue\immediate\openout\reffile=\jobname.ref%
  \def\writer@f##1>>{\ifr@ffile\immediate\write\reffile%
    {\noexpand\refis{##1} = \csname r@fnum##1\endcsname = %
     \expandafter\expandafter\expandafter\strip@t\expandafter%
     \meaning\csname r@ftext\csname r@fnum##1\endcsname\endcsname}\fi}%
  \def\strip@t##1>>{}}

\def\citeall#1{\xdef#1##1{#1{\noexpand\cite{##1}}}}
\def\cite#1{\each@rg\citer@nge{#1}}	

\def\each@rg#1#2{{\let\thecsname=#1\expandafter\first@rg#2,\end,}}
\def\first@rg#1,{\thecsname{#1}\apply@rg}	
\def\apply@rg#1,{\ifx\end#1\let\next=\relax
\else,\thecsname{#1}\let\next=\apply@rg\fi\next}

\def\citer@nge#1{\citedor@nge#1-\end-}	
\def\citer@ngeat#1\end-{#1}
\def\citedor@nge#1-#2-{\ifx\end#2\r@featspace#1 
  \else\citel@@p{#1}{#2}\citer@ngeat\fi}	
\def\citel@@p#1#2{\ifnum#1>#2{\errmessage{Reference range #1-#2\space is bad.}%
    \errhelp{If you cite a series of references by the notation M-N, then M and
    N must be integers, and N must be greater than or equal to M.}}\else%
 {\count0=#1\count1=#2\advance\count1
by1\relax\expandafter\r@fcite\the\count0,%
  \loop\advance\count0 by1\relax
    \ifnum\count0<\count1,\expandafter\r@fcite\the\count0,%
  \repeat}\fi}

\def\r@featspace#1#2 {\r@fcite#1#2,}	
\def\r@fcite#1,{\ifuncit@d{#1}
    \newr@f{#1}%
    \expandafter\gdef\csname r@ftext\number\r@fcount\endcsname%
                     {\message{Reference #1 to be supplied.}%
                      \writer@f#1>>#1 to be supplied.\par}%
 \fi%
 \csname r@fnum#1\endcsname}
\def\ifuncit@d#1{\expandafter\ifx\csname r@fnum#1\endcsname\relax}%
\def\newr@f#1{\global\advance\r@fcount by1%
    \expandafter\xdef\csname r@fnum#1\endcsname{\number\r@fcount}}

\let\r@fis=\refis			
\def\refis#1#2#3\par{\ifuncit@d{#1}
   \newr@f{#1}%
   \w@rnwrite{Reference #1=\number\r@fcount\space is not cited up to now.}\fi%
  \expandafter\gdef\csname r@ftext\csname r@fnum#1\endcsname\endcsname%
  {\writer@f#1>>#2#3\par}}

\def\ignoreuncited{
   \def\refis##1##2##3\par{\ifuncit@d{##1}%
     \else\expandafter\gdef\csname r@ftext\csname
r@fnum##1\endcsname\endcsname%
     {\writer@f##1>>##2##3\par}\fi}}

\def\r@ferr{\endreferences\errmessage{I was expecting to see
\noexpand\endreferences before now;  I have inserted it here.}}
\let\r@ferences=\references
\def\references{\r@ferences\def\endmode{\r@ferr\par\endgroup}}

\let\endr@ferences=\endreferences
\def\endreferences{\r@fcurr=0
  {\loop\ifnum\r@fcurr<\r@fcount
    \advance\r@fcurr by 1\relax\expandafter\r@fis\expandafter{\number\r@fcurr}%
    \csname r@ftext\number\r@fcurr\endcsname%
  \repeat}\gdef\r@ferr{}\global\r@fcount=0\endr@ferences}

\let\r@fend=\endpaper\gdef\endpaper{\ifr@ffile
\immediate\write16{Cross References written on []\jobname.REF.}\fi\r@fend}

\catcode`@=12

\citeall\refto		
\citeall\ref		%
\citeall\Ref		%

\referencefile
\def\uY{{U(1)_{\ssc Y}}}
\def\uYp{{U(1)_{\ssc Y^{\prime}}}}

\def\uem{{U(1)_{\ssc EM}}}
\def\suL{{SU(2)_{\ssc L}}}

\def\suc{{SU(3)_{\ss C}}}
\def\sucp{{SU(3)_{\ss {C^\prime}}}}
\def\sucpp{{SU(3)_{\ss {C^{\prime\prime}}}}}
\def\sucppp{{SU(3)_{\ss {C^{\prime\prime\prime}}}}}
\def\suPS{{SU(4)_{\ssc PS}}}
\def\suPSp{{SU(4)_{\ssc PS^\prime}}}
\def\tc{{SU(4)_{\ss TC}}}

\def\uR{{U(1)_{\ssc R}}}
\def\ssc{\scriptscriptstyle}
\def\ss{\scriptscriptstyle}
\def\nugrant{This work was supported in part by the National Science
Foundation Grants PHY-90-01439 and PHY-917809.}
\def\neuphys{Department of Physics\\Northeastern University\\Boston MA 02115}
\font\titlefont=cmr10 scaled\magstep3
\def\bigtitle                      
  {\null\vskip 3pt plus 0.2fill
   \beginlinemode \doublespace \raggedcenter \titlefont}
\oneandahalfspace
\preprintno{NUB-3058-92TH}
\preprintno{November, 1992}
\bigtitle{Self-Breaking Technicolor}
\bigskip
\author Stephen P. Martin
\affil\neuphys
\body

\abstract
We propose a scenario in which the electroweak symmetry is spontaneously
broken by an $SU(4)$ technicolor gauge interaction which also manages to
break itself completely. The technicolor gauge bosons and technifermions are
not confined by the technicolor force, but get large masses. Starting with a
single technidoublet, one emerges with a complete standard model family of
technifermions after the symmetry breaking is complete. This suggests a broad
new avenue for model building. A few variations on the theme are mentioned.

\endtitlepage
\oneandahalfspace

\subhead{1. Introduction}
\taghead{1.}

Technicolor[\cite{TC}] models of electroweak symmetry breaking are
an attempt to break the electroweak symmetry using strongly coupled gauge
theories rather than a fundamental Higgs scalar. Although technicolor does
provide a natural solution to the hierarchy and triviality problems associated
with the fundamental Higgs scalar, it is less successful at explaining
the Yukawa couplings of the standard model. The simplest attempts at
constructing a realistic fermion mass generation using extended
technicolor[\cite{ETC}] (ETC) lead to flavor-changing neutral
currents at unacceptable levels[\cite{FCNCs}]. There are other
problems associated with pseudo-Nambu-Goldstone bosons
which can be too light. Some of the proposals to avoid
such problems include the assumption of a UV fixed point for the technicolor
interaction[\cite{Holdom}], ``walking" technicolor[\cite{walking}]
in which the running of the technicolor coupling constant with scale
is taken to be much slower than in QCD,
and strong ETC models[\cite{strongETC}] in which the ETC interactions
assist in the symmetry breaking. A common idea behind these attempts is to
produce a drastic enhancement of the technifermion condensate compared to the
Nambu-Goldstone-boson decay constant. This requires that the
technicolor dynamics be very different from the dynamics of QCD.

In view of the difficulty in constructing a realistic technicolor model,
it seems worthwhile to search for qualitatively different ways to realize the
strongly coupled sector. One of the most radical qualitative changes one can
make to the original QCD-like version of technicolor is to spontaneously
break technicolor itself. This general idea has appeared recently in the
literature under various guises. Renormalizable
versions[\cite{topcolor},\cite{me},\cite{LR},\cite{KMP},\cite{King},\cite{EKR}]
of the top-quark condensate idea[\cite{Nambu},\cite{BHL}] have used a
strongly coupled and spontaneously broken interaction in which the top quark
plays the role of a techniquark. Models of spontaneously broken technicolor
have also been proposed recently by Luty[\cite{Luty}],
Sundrum[\cite{Sundrum}] and  Hill, Kennedy, Onogi and Yu[\cite{HKOY}].
The latter work considered the possibility that if technicolor is spontaneously
broken and thus does not confine, then contributions to the electroweak
$S$ parameter[\cite{oblique}] (which poses yet another phenomenological
challenge for technicolor) can be reduced somewhat. Chivukula,
Golden and Simmons[\cite{CGS}] have recently argued that the most
enormous sort of
hierarchies which have sometimes been proposed for models of this general
type will typically be ruined by a Coleman-Weinberg instability unless the
effective Higgs sector contains only one doublet.
Here, we will be mainly concerned with one paradigm for realizing
spontaneously broken technicolor and with the idea that the technicolor
interaction manages to break itself completely, leaving behind only the
standard model gauge group. We will make no attempt here to construct
a fully realistic model; in fact we will make no specific
proposal for the ETC sector.
Instead we will focus on some model building ideas as they affect the strong
coupling sector only. Our motivation is to illustrate that technicolor need
not be an unbroken, confining, QCD-like theory.

\subhead{2. Technicolor Unconfined}
\taghead{2.}

Let us choose for our gauge group
$
G= \tc \times \sucp \times \suL \times \uYp .
$
There is one doublet of technifermions which transforms under $G$ as
$$
\eqalign{
& \psi_{\ssc L} \cr
& \psi^c_{\ssc R} \cr
}
\eqalign{
\> \sim \> & ({\bf 4}, {\bf 1}, {\bf 2}, 0)    \cr
\> \sim \> & ({\bf {\overline 4}}, {\bf 1}, {\bf 1}, { 1/2})
+ ({\bf {\overline 4}}, {\bf 1}, {\bf 1}, { -1/2}) \>\> . \cr
}
\eqno(technis)
$$
[The gauge transformation properties of fermions are always given in
terms of left-handed two-component Weyl fields in \(technis) and throughout
the rest of this paper.] The standard model quarks and leptons transform
under $G$ as three copies of
$$
\eqalign{
& (u\, d)_{\ssc L} \cr
& (\nu\, e)_{\ssc L} \cr
}
\eqalign{
\> \sim \>  &({\bf 1}, {\bf 3}, {\bf 2}, { 1/6})    \cr
\> \sim \>  &({\bf 1}, {\bf 1}, {\bf 2}, { -1/2})    \cr
}
\qquad\>\>\>
\eqalign{
& d^c_{\ssc R} \cr
& e^c_{\ssc R} \cr
}
\eqalign{
\> \sim \>  &({\bf 1}, {\bf {\overline 3}}, {\bf 1}, { 1/3}) \cr
\> \sim \>  &({\bf 1}, {\bf 1}, {\bf 1}, 1) \>\> .
\cr}
\qquad\>\>\>
\eqalign{
& u^c_{\ssc R} \cr
&   \cr }
\eqalign{
\> \sim \>  &({\bf 1}, {\bf {\overline 3}}, {\bf 1}, { -2/3}) \cr
& \cr}
\eqno(ql)
$$
Since $\tc$ is asymptotically free, it will get strong in the infrared.
The electroweak symmetry is then broken by the condensate
$$
\langle \overline{\psi}_{\ss L}^i {\psi_{\ss R}}_j \rangle = \mu^3 \delta^i_j
\qquad\qquad i,j = 1,2
\eqno(con)
$$
as usual in technicolor models.

So far this is just the one-doublet $SU(4)$ technicolor model.
But now we consider the possibility that the technicolor gauge group
itself is broken. Specifically, suppose that
$ \tc \times \sucp \times \uYp$ is broken down to the diagonal
$\suc \times \uY$. [More precisely, this means that $\suc$ is the diagonal
subgroup of $SU(3)_{\ssc TC} \times \sucp$ and $\uY$ is the diagonal subgroup
of $U(1)_{\ssc TC}$ and $\uYp$, where $SU(3)_{\ssc TC} \times U(1)_{\ssc TC}$
is a maximal proper subgroup of $\tc$.] The unbroken
$\suc$ and $\uY$ are the standard model color and weak hypercharge,
respectively. The scale $M$ which characterizes the breaking
$ \tc \times \sucp \times \uYp \rightarrow \suc \times \uY$ should not be
too large, otherwise $\tc$ will get broken before it has a chance to
get strong enough to form the condensate \(con).

Of the twenty-four gauge bosons associated with
$ \tc \times \sucp \times \uYp$, eight gluons and one hyperphoton remain
massless, corresponding to the unbroken gauge group $\suc \times \uY$.
 At the scale
$M$, the coupling constants for the unbroken gauge groups are given by
$$
g_3 = {{g_{3^\prime} g_4} / \sqrt{g_4^2+ g_{3^\prime}^2}}
\qquad {\rm and} \qquad
g_1 = g_{1^\prime} g_4 / \sqrt{g_4^2+ 2 g_{1^\prime}^2/3}
\eqno(couplings)
$$
where $g_{3^\prime}$ and $g_{1^\prime}$ are the coupling constants for
$\sucp$ and $\uYp$ respectively.
Now we are assuming that $\tc$ is strongly coupled at $M$ and that $\sucp$,
$\suL$ and $\uYp$ are not, so that $g_4 \gg g_{3^\prime}, g_{2}, g_{1^\prime}$.
Therefore $g_3 \approx g_{3^\prime}$ and $g_1 \approx g_{1^\prime}$ at
the scale $M$.
The other fifteen gauge bosons, consisting of a color octet,
triplet and antitriplet, and singlet, all get masses of
order $g_4 M$, where $g_4$ is the coupling constant for $\tc$.

After integrating out the fifteen heavy gauge bosons, one finds the
following four-fermion interaction (to lowest order in $g_{3^\prime}/g_4$
and $g_{1^\prime}/g_4$):
$$
L_{\rm eff} = -{g_4^2 \over 2M^2} J_\mu^A J_A^{\dagger\mu};\qquad\qquad
J_\mu^A =
{\overline\psi}_{\ss L}^i \gamma_\mu T^A \psi_{{\ssc L}i} +
 {\overline\psi}_{\ssc R}^i \gamma_\mu T^A \psi_{{\ssc R}i}
\>\> .
\eqno(4f)
$$
[Here $T^A$ are the fifteen generators of $SU(4)$.] This
interaction can be Fierzed into a form which includes the term
$$
L_{\rm eff} = {15\over 16}{ g_4^2 \over M^2}
({\overline\psi}_{\ssc L}^i \psi_{{\ssc R}j})
({\overline\psi}_{\ssc R}^j \psi_{{\ssc L}i})
\>\> .
\eqno(fierzed)
$$
This term is just a Nambu--Jona-Lasinio[\cite{NJL}] (NJL)
interaction representing an attractive force between the technifermions.
One might now formulate a NJL interpretation of the strong
coupling dynamics, in which if $g_4$ is sufficiently
large at the scale $M$, then the condensate \(con) will form. This
is a ruthlessly truncated version of the symmetry breaking dynamics,
in which everything except technigluon exchange at zero momentum transfer
(and in the $\tc$-singlet channel after Fierzing) is
neglected. It therefore seems prudent to refrain from attempting
to use a NJL analysis to obtain a quantitative description of the
symmetry breaking. Still, the NJL picture can provide a useful qualitative
picture. The idea presented here can then be
viewed as a specific renormalizable realization of refs. [\cite{HLP}] and
[\cite{HKOY}].

After the symmetry breaking
$ \tc \times \sucp \times \uYp \rightarrow \suc \times \uY$,
the technidoublet $\psi_{\ss L} , \psi_{\ss R}^c$ transforms as one complete
standard model family (including a gauge singlet technineutrino)
under $\suc \times \suL \times \uY$:
$$
\eqalign{
& \psi_{\ss L} \cr
& \psi^c_{\ss R} \cr
}
\eqalign{
\> \sim \> & ({\bf 3}, {\bf 2}, { 1/6})
+ ({\bf 1}, {\bf 2}, { -1/2})   \cr
\> \sim \> & ({\bf \overline 3}, {\bf 1}, { -2/3})
+ ({\bf 1}, {\bf 1}, 0)
+ ({\bf \overline 3}, {\bf 1}, { 1/3}) + ({\bf 1}, {\bf 1}, 1)
\>\> . \cr }
\eqno(btechnis)
$$
These technifermions all get large masses associated with the
electroweak-breaking condensate \(con). The standard model quarks and leptons
transform in exactly the same way, (except that there is no gauge singlet
neutrino). From the point of view of low energy physics, $\tc$ is a
Pati-Salam symmetry for the techniquarks only. The heavy octet and heavy
singlet gauge bosons also couple to the standard model fermions,
albeit weakly. This is because the mass eigenstates
are related to the original gauge eigenstates of the vector bosons by a
small mixing angle,  with the heavy vectors consisting mostly
of $\tc$ gauge bosons which do not couple to the standard model
quarks and leptons. The heavy color octet couplings to quarks are
$\approx g_3^2/g_4$ and exactly
mimic those of a heavy gluon, and the heavy gauge singlet couplings
to quarks and leptons are $\approx \sqrt{2/3} g_1^2/g_4$ times the
weak hypercharge. The heavy color triplet and antitriplet gauge bosons have
no direct couplings at all to the standard model quarks and leptons.

Some four-fermion effective interactions presumeably must be included to
communicate the electroweak breaking to the standard model quarks and
leptons and provide for their masses. These should have the schematic form
$ {(\lambda^2 / \Lambda_{\ss ETC}^2)}
({\overline q}_{\ssc L} q_{\ssc R})({\overline \psi}_{\ssc L} \psi_{\ssc R})
+ {\rm h.c.}$ or
$ {(\lambda^2 / \Lambda_{\ss ETC}^2)}
({\overline q}_{\ssc R} q_{\ssc L})({\overline \psi}_{\ssc L} \psi_{\ssc R})
+ {\rm h.c.}$ for the quarks, and similarly for the charged leptons.
We will not speculate in this paper on what underlying dynamics gives rise to
such interactions.

How can we break $\tc$?
One possibility is to simply introduce a scalar field which gets
a vacuum expectation value. For example, one might introduce a scalar
$\Phi$ in the rep $ ({\bf 4}, {\bf \overline 3}, {\bf 1}, -1/6)$ of $G$.
Then if $\Phi$ develops a vacuum expectation value
$$
\langle \Phi^a_\alpha \rangle = M \delta^a_\alpha \qquad\qquad
a=1,2,3,4; \> \alpha = 1,2,3
\eqno(Phi)
$$
the symmetry will break in the desired way. Of course, if $\Phi$ is a
fundamental scalar, then it carries with it the usual hierarchy and
triviality problems which are the main motivations for considering
technicolor models in the first place! We would then
require an explanation for why loop corrections do not push $M$ to
the Planck mass or some other scale much higher than the scale at which $g_4$
becomes large enough to admit the possibility of condensates.
Therefore, we prefer to consider the possibility that $\Phi$, or some
analogous order parameter(s), is itself a composite
field corresponding to a condensate of fermions in a scalarless theory.

\subhead{3. A Paradigm for Self-Breaking}
\taghead{3.}

An economical possibility is that $\tc$ manages to break itself without the
introduction of an additional strongly coupled interaction.
Suppose there are additional fermions
which transform nontrivially under both $\tc$ and  $\sucp \times \uYp$,
and that they can be chosen in such a way  that they also form condensates
due to the strong $\tc$ force which then break $\tc$ in the right way.
It is clear that we will need at least one fermion which transforms in
a higher dimensional rep of $\tc$.
For, if only $\bf 4$s and $\bf \overline 4$s are present, then in order
for the $\tc$ anomaly to cancel, the number of $\bf 4$s must equal the
number of $\bf \overline 4$s. As long as this is the case, then the
condensates will always occur as $\tc $ singlets (because the $\bf 4$s will
just pair up with the ${\bf{\overline 4}}$s) and therefore cannot
break $\tc$. On the other hand, if there are too many
fermions in higher dimensional reps
of $\tc$ then the $\tc$ $\beta$ function will be positive.
This would be a disaster since we need $g_4$ to get large
in the infrared. This is a welcome restriction which relieves us of the
responsibility of considering models which are too baroque.

Here is one scenario for self-breaking of $\tc$. In addition to
the technifermions in \(technis) and the standard model quarks and leptons
in \(ql), we assign fermions to the following reps of $G$:
$$
\eqalign{
\eta \> \sim \>  ({\bf 10}, {\bf 1}, {\bf 1}, 0);    \qquad\qquad\>
\xi \> \sim \>  ({\bf \overline 4}, {\bf \overline 3}, {\bf 1}, -1/6) & +
({\bf \overline 4}, {\bf {1}}, {\bf 1}, 1/2);
\cr
\chi \> \sim \>  ({\bf \overline 4}, {\bf 3}, {\bf 1}, 1/6)  +
({\bf \overline 4}, {\bf {1}}, {\bf 1}, -1/2) & \>\> .  \cr
}
\eqno(newtechnis)
$$
[Note that the choice of which
$({\bf \overline 4}, {\bf 1}, {\bf 1}, { \pm 1/2})$ belong
to $\psi_{\ssc R}^c$ and which to $\xi, \chi$ is just an arbitrary choice of
orientation.]   All of the gauge anomalies cancel for the fermions in
\(newtechnis). We have grouped the new fermions in the way indicated
in \(newtechnis) for the following reason. The gauge group
$\sucp \times \uYp$ can be embedded into an approximate global symmetry
group $\suPSp \times \uR$ with $\suPSp$ an ungauged Pati-Salam symmetry.
Now all of the fermions can be arranged into multiplets which are
irreducible reps of $\tc \times \suPSp \times \suL \times \uR$.
Under this group, $\eta^{(ab)}$ transforms as
$({\bf 10}, {\bf 1}, {\bf 1}, 0)$, $\xi_{a\alpha}$ transforms as
$({\bf \overline 4}, {\bf \overline 4}, {\bf 1}, 0)$,
and $\chi^\alpha_a$ as $({\bf \overline 4}, {\bf 4}, {\bf 1}, 0)$.
[Latin letters $a,b,c\ldots$
represent $\tc$ indices in the fundamental rep and Greek letters
$\alpha,\beta,\gamma$ represent $\suPSp$ indices, with
$\alpha,\beta,\gamma = 1,2,3$ corresponding
to the gauged subgroup $\sucp$. Note that $\eta^{(ab)}$
carries two symmetrized
indices since the $\bf 10$ of $SU(4)$ is the symmetrized direct
product of two $\bf 4$s.] The technifermions $\psi^a_{\ssc L}$ and
$\psi_{{\ss R} a}^c$ transform as
$({\bf 4},{\bf 1},{\bf 2},0)$ and
$({\bf \overline 4}, {\bf 1}, {\bf 1}, \pm 1/2)$ and the standard model
fermions as $({\bf 1},{\bf 4},{\bf 2},0)$ and
$({\bf 1},{\bf \overline 4},  {\bf 1}, \pm 1/2)$. This classification
is particularly useful because it will allow us to write the condensates
which can occur in this model in a compact notation.
[$\suPSp$ may also be a gauge symmetry broken at some energy scale
much higher than any other scale of interest in this paper.]

Now $\tc$ is still asymptotically free;
its $\beta$ function is given to two loops by
$$
\beta_4 = \mu {d g_4 \over d\mu} = -{g_4^3 \over 16 \pi^2} \left [
{26\over 3} + {71\over 6} \left ({g_4\over4 \pi}\right )^2  \right ]
\>\> .
$$
So the coupling constant $g_4$ grows in the infrared.
When $\tc$ gets strong enough, condensates involving $\psi_{\ssc L}$,
$\psi^c_{\ssc R}$, $\eta$, $\xi$, and $\chi$ should form. A rigorous
discussion of this process would require a complete understanding of the
non-perturbative dynamics of strongly coupled and spontaneously
broken gauge interactions, which we do not have.
In order to make progress, we must simply choose a plausible set of
assumptions about the strong coupling dynamics and hope that they are correct.

To help decide qualitatively how the condensates arrange themselves, we can
make use of the single gauge boson approximation, which we now briefly review.
Consider a model which consists of an
asymptotically free gauge theory [in our case $\tc$] which couples to
some fermions but no scalars. The fermions may also have weakly coupled
gauge interactions [in our case $\sucp \times \suL \times \uYp$]
whose effects may be treated perturbatively.
When the strong gauge coupling becomes sufficiently large in the infrared,
scalar fermion bilinear condensates will form in irreducible
reps of the gauge group. Suppose that the fermions involved
in the condensate transform under the strongly coupled gauge group in the
irreducible reps $R_1$ and $R_2$, and the resulting
condensate transforms as $R_s$.
Thus $R_s$ occurs in the direct sum decomposition of the direct
product $R_1 \times R_2 = R_s + \cdots$.
We need a way of deciding for which choices of $R_1$, $R_2$, and $R_s$
the condensate will occur. According to the single gauge boson exchange
approximation, the condensates will tend to appear in
the ``most attractive scalar channel" (MASC), $R_1 \times R_2 \rightarrow R_s$,
for which $V=  C_1 + C_2 - C_s$ is largest. Here $C_1$, $C_2$, and $C_s$
are the quadratic Casimir invariants for the representations
$R_1$, $R_2$, and $R_s$, respectively.
When a given fermion condenses, it obtains a self-energy term
which suppresses its ability to participate in other condensates.
We therefore assume that each fermion condenses at most once.

In the case of our model, the strongly coupled $\tc$ has fermions
transforming as a $\bf 10$, ten $\bf \overline 4$s, and two $\bf 4$s.
The attractive channels for this fermion content, and their relative
strengths $V$ are as follows\footnote{$^\dagger$}{All group theory
conventions and facts used in this paper may be found from [\cite{Slansky}].}:
$$
\eqalign{
& {\underline {\rm Channel}} \cr
(i) \qquad   &{ \bf 10} \times  {\bf \overline 4} \rightarrow { \bf 4} \cr
(ii) \qquad   &{ \bf 4} \times  {\bf\overline 4} \rightarrow { \bf 1}  \cr
(iii) \qquad  &{ \bf 10} \times  {\bf 10} \rightarrow { \bf 20^\prime} \cr
(iv) \qquad  &{ \bf 10} \times  {\bf 4} \rightarrow {\bf {\overline {20}}} \cr
}
\qquad
\eqalign{
& {\underline V} \cr
& 18 \cr
& 15 \cr
& 12 \cr
& 6 \cr
}
\qquad\qquad\>\>
\eqalign{
& {\underline {\rm Channel}} \cr
(v) \qquad   &{\bf\overline 4}\times {\bf \overline 4} \rightarrow {\bf 6} \cr
(vi) \qquad   &{\bf 4}\times {\bf 4} \rightarrow { \bf 6}  \cr
(vii) \qquad  &{ \bf 10} \times  {\bf 10} \rightarrow { \bf 45} \cr
& {} \cr
}
\qquad
\eqalign{
& {\underline V} \cr
& 5 \cr
& 5 \cr
& 4 \>\> . \cr
& {} \cr
}
$$
Since channel $(i)$ is the MASC in this naive approximation,
we assume that $\eta$ condenses with some combination of
$\bf \overline 4$s ($\xi$, $\chi$ and $\psi^c_{\ss R}$).

However, the MASC criterion still leaves an important
ambiguity, since the $\bf 10$ can condense with one or with
more than one $\bf \overline 4$. Indeed, there is always such an ambiguity
in the MASC criterion whenever one of the reps appearing in the
MASC occurs more than once in the list of massless fermions.
To understand this ambiguity more clearly, let us use a notation in which
all ten of the $\bf \overline 4$s are represented by a generic symbol
$f_a^{\ssc I}$ with $I=1\ldots 10$. Then the fact that the condensate occurs
in channel $(i)$ is precisely equivalent to the statement that it has the form
$$
\langle \eta^{(ab)} f_c^{\ss I} \rangle = \delta^{(a}_c v^{b)}_{\ss I}
M^3
\eqno(ambi)
$$
where $v^a_{\ss I}$ is unknown at this point, and determines how much
of the $\tc$ and global symmetries are broken. The determination
of $v^a_{\ss I}$ is {\it not} just a vacuum alignment problem having
to do with residual gauge symmetries; it is properly
to be determined by the strongly coupled part of the theory.
The question of how to resolve such ambiguities has been addressed by
Gusynin, Miransky, and Sitenko[\cite{GMS}]. Using their arguments, based
on a stability analysis within a solvable approximation,
we find (up to global $SU(10)$ and $\tc$ rotations)
$$
v^a_{\ss I} = \delta^a_{\ss I} \qquad (I=1\ldots 4);
\>\>\qquad\qquad v^a_{\ss I} = 0 \qquad (I=5\ldots 10) \>\> .
\eqno(answer)
$$
This agrees with the heuristic criterion in [\cite{GMS}] that when
such an ambiguity exists, the number of fermions which condense in the
MASC is maximized. (See example (a) in [\cite{GMS}] for a problem which
is exactly analogous to the one discussed here.)
Now, \(answer) means that $\tc$ is completely broken in one step.
It is not broken to a strongly coupled
$SU(3)$ subgroup, which would require instead
$v^a_{\ss I} = \delta^a_4 \delta^1_{\ss I}$.
Of the original symmetry $\tc\times SU(10)$, a global $SU(4) \times SU(6)$
is left unbroken by \(answer).

With $v^a_{\ss I}$ given by \(answer), there still remains a
vacuum alignment problem, having to do with the orientation
of the weakly coupled gauge group $\sucp \times\suL \times \uYp$
with respect to the surviving global symmetry.
This type of vacuum alignment problem has been studied in
[\cite{alignment}].
The vacuum tends to align so as to preserve as much of the residual gauge
symmetry as possible. There turn out to be two distinct and equally
good solutions to this vacuum alignment problem, one in which
$\eta$ condenses entirely with $\chi$ and one in which it condenses
entirely with $\xi$. We imagine for now that unspecified higher order
effects prefer the latter solution, and will discuss the situation
if the former solution wins in the next section.
Thus $\eta$ condenses with $\xi$ according to
$$
\langle \eta^{(ab)} \xi_{c\alpha} \rangle =
\delta^{(a}_c \delta^{b)}_\alpha   \> M_{\eta\xi}^3
\>\> .
\eqno(con2)
$$
There is a simple heuristic reason for the vacuum alignment
\(con2); if $\suc$ is left unbroken,
then the fermion pairs participating in the condensates will feel an additional
attractive force due to $QCD$, because they transform as conjugate
representations of
$\suc$. Thus the condensates will align as in \(con2) so as not to break
$\suc$, because they can. The condensate \(con2) transforms under
$\tc \times \suPSp$
as $({\bf 4}, {\bf \overline 4})$, and breaks $\tc \times \suPSp$
down to the diagonal $\suPS$ of the standard model. [This is easily seen
because the right-hand side of \(con2) is an invariant symbol of
$\suPS$.]
Thus it also breaks the gauged subgroup $\tc \times \sucp \times \uYp$
down to $\suc \times \uY$ as desired.

Now we make an additional assumption. Since channel $(ii)$
is only weaker than channel $(i)$ by a factor of $6/5$ in the single
gauge boson exchange approximation, we assume that channel
$(ii)$ also condenses.
We assume that this is true even though the condensate of channel $(i)$
breaks the $\tc$ interaction. This assumption is equivalent to the assumption
in the NJL language that the four-fermion interaction \(4f) is sufficiently
strongly coupled to produce the condensate \(con). Now, there is again
a vacuum alignment problem, since the two $\bf 4$s of $\psi_{\ss L}$ have
a choice of $\bf \overline 4$s with which to condense, namely
$\psi_{\ss R}^c$, $\chi$, and the antisymmetric part
of $\xi$ ($\xi_{\ss A}$) which did not
condense with $\eta$. Again, the vacuum chooses to align itself so
that $\suc$ and $\uem$ are unbroken, because that allows for an additional
attractive force between the condensing fermion pairs. That is why
$\psi_{\ss L}$ condenses with $\psi^c_{\ss R}$ as in \(con), and not in some
other way which would break $\suc$ or $\uem$.

To recapitulate, we are assuming that the two condensates \(con)
and \(con2) both form, with roughly equal strength. The condensate \(con)
breaks the electroweak symmetry and the condensate \(con2) breaks
technicolor.  The scalar $\Phi$ of section 2 is replaced by
the condensate $\langle \eta \xi \rangle$.

After the condensate \(con2) forms, the fermions $\eta$, $\xi$ and $\chi$
transform under the standard model gauge group $\suc \times \suL \times \uY$ as
$$
\eqalign{
& \eta
\> \sim \> ({\bf 6}, {\bf 1}, 1/3) + ({\bf 3}, {\bf 1}, -1/3) +
({\bf 1}, {\bf 1}, -1)
\cr &
\xi_{\ssc S}
\> \sim \> ({\bf \overline 6}, {\bf 1}, -1/3)
+ ({\bf \overline 3}, {\bf 1}, 1/3) + ({\bf 1}, {\bf 1}, 1)  \cr &
\xi_{\ssc A}
\> \sim \> ({\bf \overline 3}, {\bf 1}, 1/3) + ({\bf 3}, {\bf 1}, -1/3)  \cr &
\chi
\> \sim \> ({\bf 8}, {\bf 1}, 0) + ({\bf \overline 3}, {\bf 1}, -2/3) +
({\bf 3}, {\bf 1}, 2/3) + 2 \times ({\bf 1}, {\bf 1}, 0)
\>\> .
\cr
}
\eqno(breakdown)
$$
This shows that the QCD force aligning the condensates can be quite
appreciable, especially since a color sextet from $\eta$ and an antisextet from
$\xi_{\ssc S}$
are paired in the condensate \(con2). This should firmly stabilize the
$\suc$--conserving vacuum. At this point, $\eta$ and $\xi_{\ssc S}$
pair up and obtain an effective mass term.
So far, $\xi_{\ssc A}$ and $\chi$ remain uncondensed and massless,
even though they transform as a real representation of the standard
model gauge group. Since these fermions have not been seen yet in nature,
we must specify a mechanism to obtain large mass terms
$\sim m_{\xi_{\ssc A}} \xi_{\ssc A} \xi_{\ssc A}$ and
$\sim m_{\chi} \chi \chi$.
These mass terms are allowed by
the standard model gauge group, but they are not
generated at this stage because they break additional chiral
symmetries which are left unbroken by \(con) and \(con2).

A traditional assumption of ``tumbling" gauge theories[\cite{tumbling}]
is that when a condensate breaks the strongly coupled gauge theory,
then any condensate corresponding to a channel which is weaker in
the single gauge boson exchange approximation will not form. The idea
behind this is that when the strongly coupled gauge bosons gain mass,
they decouple and so their ability to produce condensates is vitiated.
We have already assumed that this is not quite the case,
since we assumed that condensates occur in both channels $(i)$ and
$(ii)$, even though channel $(i)$ is naively slightly stronger and breaks
$\tc$. Qualitatively, this is an assumption that the mass of the gauge boson
is not large enough to prevent the condensate in channel $(ii)$.
One way for $\xi_{\ss A}$ and $\chi$ to get large masses uses
a slightly untraditional (but not, we think, outrageous) set of assumptions
as follows. First, we continue to assume that each fermion can participate
in at most one condensate. The rationale behind this is that once a fermion
condenses, its self-energy so generated inhibits its ability to participate
in further condensates. However, we now consider the further assumption
that each strongly coupled fermion does condense exactly once. This could
come about if the mass obtained by  the strongly coupled gauge bosons is
small enough to allow condensates to form in all attractive channels
involving massless fermions.

Then the
analysis of the condensation pattern might go as follows. As before,
the condensates \(con) and \(con2) appear due to channels $(ii)$ and $(i)$
respectively. Now because each fermion can condense only once (by assumption),
we may forget about channels $(iii)$, $(iv)$, $(vi)$ and $(vii)$, since all
of the $\bf 10$s and $\bf 4$s of $\tc$ have been used up. The remaining
fermions thus condense according to channel $(v)$, even though this channel
is weaker than channels $(i)$ and $(ii)$. The resulting
condensates will take the form:
$$
\eqalignno{
&\langle \xi_{a\alpha} \xi_{b\beta} \rangle =
\epsilon_{a\alpha b\beta}   \> M_{\xi\xi}^3
&(con3)
\cr
&\langle \chi^\alpha_a \chi^\beta_b \rangle =
\delta^{[\alpha}_a \delta^{\beta ]}_b      \> M_{\chi\chi}^3
\>\> .
&(con4)
\cr
}
$$
(The brackets in \(con4) indicate antisymmetrized indices.)
Once again we have used the fact that these condensates will align so as
to leave $\suc$ unbroken. Each of the condensates $\langle \xi \xi\rangle$
and $\langle\chi\chi\rangle$ transforms under $\tc \times \suPSp$ as
$({\bf 6}, {\bf 6})$, and leaves the diagonal $\suPS$ with its gauged
subgroup $\suc \times \uY$ unbroken, since the right-hand sides
of \(con3) and \(con4) are again invariant symbols of $\suPS$.
One component of $\langle \chi\chi\rangle$ involves a QCD octet
condensing with itself. This stabilizes the $\suc$--conserving vacuum
and also can enhance the condensate considerably.

Note that now,
each of the components of $\eta,\xi,\chi$ participates in exactly one
condensate. The symmetric component of $\xi$ condenses with $\eta$ and
the antisymmetric component of $\xi$ condenses with itself.
Modulo the assumptions stated above, this is one way of assuring that all
of the fermions $\eta$, $\xi$ and $\chi$ obtain large masses.
In the next section we will discuss a variation of the basic model
in which $\xi_{\ssc A}$ and $\chi$ get their masses from a
strongly coupled interaction which also produces a mass for the top quark.

As usual in technicolor models, the strong interactions have a large
approximate global symmetry  which contains the electroweak symmetry as
a subgroup. A host of pseudo-Nambu-Goldstone bosons (PNGBs) arise when
the condensates break this approximate global symmetry. In our case,
the approximate symmetry of the strong interactions is
$\tc \times SU(10) \times \suL \times U(1) \times U(1)$.
(There would be another
$U(1)$ global symmetry but it is removed by instanton effects.) The condensates
\(con), \(con2), \(con3) and \(con4)
will break this down to the vectorial subgroup
$\suPS \times SU(2)_{\ssc V} \times U(1)_{\ss TB}$, which contains the
gauged subgroup $\suc \times \uem$. Here $SU(2)_{\ss V}$ is the vector-like
``custodial"[\cite{custodial}] symmetry and
$U(1)_{\ss TB}$ is a technibaryon number (which is exactly conserved
except for $\suL$ instanton effects). Thus there are 100 PNGBs in this
model. Of these, 15 are eaten when $\tc$ gets broken, and 3 more are eaten
by the $W$ and $Z$ when the electroweak symmetry is broken. The remaining
82 PNGBs transform as 14 color singlets, 16 color triplets, 2
color sextets, and one color octet. The colored PNGBs get
large masses as usual in technicolor models, while the color singlet
PNGBs presumeably get  masses from ETC and other unspecified interactions which
explicitly break the approximate global symmetry. There are also
heavy technimesons and technibaryons in this model. The lightest of
the latter should be stable because of the conservation of technibaryon number.

\subhead{4. Variations on the Theme}
\taghead{4.}

In the preceding section we discussed one way of breaking
$\tc \times \sucp \times \uYp \rightarrow \suc \times \uY$,
using a specific set of extra fermions and a specific
set of assumptions about how they behave when $\tc $ gets strong.
There is no particular reason to believe that this version is unique.
Indeed, it is not even complete, since we have not yet specified
exactly how mass terms arise for the standard model
fermions. One can imagine a multitude of variations on the theme
in which e.g.~$\eta$, $\xi$ and $\chi$ are replaced by other extra fermions,
or in which the assumptions about the behavior of the strongly coupled
interaction are modified, or in which additional gauge interactions
play a key
role in the condensate formation, etc. In this section we will mention
a few such variations on the general theme described in the previous
sections. In doing so, we will ignore the important problems of
mass generation for standard model quarks and leptons, flavor-changing
neutral currents, the fate of the PNGBs, and precision electroweak
parameters. Instead, we will content ourselves with some gross features
of the strongly coupled sector which serve to illustrate the richness of
model building possibilities.

\noindent{\it{Variation 1: Additional technidoublets and the $\beta$
function.}}
The $\beta$ function for the model described in Section 3 has healthy
negative contributions from both one and two loops. We might want to
add in more fermions which transform under $\tc$ in order to make
$g_4$ walk more slowly into the infrared. This can be done in several
ways without altering the symmetry breaking pattern. However, there
is little room to add in more strongly coupled fermions without
endangering the growth of $g_4$ in the infrared altogether.
In general, the $\beta$ function for $\tc$ is given to two loops by
$$
\eqalign{
\beta_4 & = \mu {d g_4 \over d\mu} = -{g_4^3 \over 16 \pi^2} \left [
b_0 + b_1 \left ({g_4\over4 \pi}\right )^2 \right ] \cr
b_0 & = (44 - n_4 - 2 \, n_6 - 6 \, n_{10})/ 3   \cr
b_1 & = (4352 - 205 \, n_4 - 440\, n_6 - 1608\, n_{10})/24  \cr
}
$$
where $n_4$, $n_6$ and $n_{10}$ are the total number of two-component
Weyl fermions
transforming as $\bf 4$ or $\bf \overline 4$, $\bf 6$, and $\bf 10$ or
$\bf \overline {10}$, respectively.
By using more than one technidoublet, we can slow the running of $g_4$.
For example, if we add to the model of Section 3 one extra copy of
$\psi_{\ss L}$ and $\psi_{\ss R}^c$ so that we have two technidoublets in all,
then we have $n_4 = 16$, $n_6 = 0$,
and $n_{10} = 1$, giving $b_0 = 22/3$ and $b_1 = -67/3$. The positive
two-loop contribution to the $\beta$ function overpowers the negative
one-loop contribution at $g_4/4\pi \approx .57$, which is therefore
an infrared fixed point of the two-loop $\beta$ function. Of course,
as the coupling approaches this value, the perturbative expression
for the $\beta$ function becomes untrustworthy. Still, one may
speculate that there is a fixed point for the exact theory somewhere
roughly in the vicinity of this point, or at least a very slow running of
$g_4$. If we use three
technidoublets, so that $n_4 = 20$, $n_6=0$, $n_{10} =1$, we get
$b_0 = 6$ and $b_1 = -113/2$.
Now the naive estimate for the possible fixed point is
$g_4/4 \pi \approx .33$.
Such values for the coupling may
be just big enough to give chiral symmetry breaking near criticality.
Adding in extra fermions in $\bf 6$, $\bf 10$ or larger reps of $\tc$ in
a way consistent with the desired symmetry breaking pattern tends
to give large negative contributions to $b_0$ and $b_1$, making it
problematical for $g_4$ to obtain large enough values for chiral symmetry
breaking at all.

\noindent{\it{Variation 2: Replacing $\langle \eta \xi\rangle$ with
$\langle \eta \chi \rangle$.}}
As we mentioned in the previous section, \(con2) is not the unique
solution to the vacuum alignment problem concerning the relative
orientation of the weakly coupled gauge group and the
unbroken global symmetry. The other solution is given by
$$
\langle \eta^{(ab)} \chi_{c}^{\alpha} \rangle =
\delta^{(a}_c \delta^{b)}_\alpha   \> M_{\eta\chi}^3
\>\> .
\eqno(alternate)
$$
If the theory chooses \(alternate) instead of \(con2), then the analysis
we have already presented goes through in much the same way, by
interchanging the roles of $\chi$ and $\xi$
and applying the conjugation automorphism to $\tc$.
If $SU(4)^*_{\ss TC}$ represents the same group as $\tc$ but with each
representation replaced by its conjugate, then the condensate
\(alternate) transforms under
$SU(4)^*_{\ss TC}\times\suPSp$ as $({\bf\overline 4},{\bf 4})$
and breaks $SU(4)^*_{\ss TC}\times\sucp \times \uYp$ down to
$\suc\times\uY$. The net effect of this is that the representations
of the technifermions under the low energy gauge group are all replaced
by their conjugates. The fermions $\eta$, $\xi$, and $\chi$
together form a real representation of the standard model gauge group
and their low-energy quantum numbers are unaffected.
However, the technifermions $\psi_{\ss L}$ and $\psi^c_{\ss R}$ transform
under the standard model gauge group as the conjugates of the reps in
\(btechnis) in this variation.
This will make a difference in trying to construct ETC interactions.

\noindent{\it{Variation 3: A hybrid technicolor and top-quark
condensate model.}}
The ``zeroeth order" spectrum of standard model fermion masses consists
of a large top-quark mass and negligible masses for everything else.
Let us now briefly sketch a model which exhibits this spectrum, and
which incidentally ensures large masses for $\xi$ and $\chi$. It is
really just a hybrid of the spontaneously broken technicolor
model in sections 2 and 3 and a renormalizable top-quark condensate
model of the ``Topcolor"[\cite{topcolor},\cite{me}] type.
The full gauge group is
$\tc\times \sucp \times \sucpp \times \suL \times \uYp$.
The fermions $\psi_{\ssc L}, \psi_{\ssc R}^c, \eta, \xi$ and $\chi$
are all singlets
with respect to $\sucpp$ and they transform under $\tc\times\sucp\times
\suL\times\uYp$ exactly as in \(technis) and \(newtechnis). As before,
we assume that $\tc$ gets strong first in the infrared. Then, exactly
as discussed in sections 2 and 3, condensates \(con) and \(con2) will
form, breaking $\tc\times\sucp\times\uYp \rightarrow \sucppp \times \uY$.
But now, instead of identifying $\sucppp$ with standard model QCD, we assume
that $\sucppp$ also gets strong above the electroweak scale. When this happens,
condensates involving the components of $\xi_{\ssc A}$ and $\chi$ will
form. Since $\xi_{\ssc A}$ and $\chi$ form a vector-like representation
of the remaining gauge group, they condense in the obvious way
without breaking any additional gauge symmetries. That is, the
$({\bf 8},{\bf 1}, 0)$ of $\chi$ condenses with itself, and
the $({\bf 3},{\bf 1},{ 2/3})$  and
$({\bf \overline 3},{\bf 1},{ -2/3})$ of $\chi$ condense with
each other, while the
$({\bf 3},{\bf 1},{ -1/3})$
and $({\bf \overline 3},{\bf 1},{ 1/3})$ of $\xi$ condense with
each other. The two copies of $({\bf 1},{\bf 1}, 0)$ in $\chi$ do not condense
due to $\sucppp$, but they still get large masses communicated to them
from the other condensates by the strongly coupled $\tc$ interactions.
Note that the $\tc$ interactions can also strongly enhance the other
condensates and mass terms, (and ensure that the $\bf 3$ and $\bf\overline 3$
components of $\xi_{\ssc A}$
and $\chi$ are not tempted to condense with the fermions we are about
to introduce).

Now $\sucppp$ plays the role of the stronger $SU(3)$ and $\sucpp$
the role of the weaker one in the ``Topcolor" paradigm.
To be specific, we can follow [\cite{me}] and
complete this model by putting in  $\tc$-singlet fermions, consisting
of the standard model quarks and leptons and two $\suL$-singlet
quixes (sextet quarks). They transform under
$\sucp\times\sucpp\times\suL\times\uYp$ as
$$
\eqalign{
& {\overline q}_1\cr
& q_1, q_2, d^c_{1\ssc R}, d^c_{2\ssc R} \cr
& {\overline q}_2 \cr
& (t \, b)_{\ssc L} \cr
& t^c_{\ssc R} \cr
}
\eqalign{
 \> \sim \> &          \cr
 \> \sim \> &          \cr
 \> \sim \> &          \cr
 \> \sim \> &          \cr
 \> \sim \> &          \cr
}
\eqalign{
         &({\bf {\overline 6}}, {\bf 1}, {\bf 1}, -1/3) \cr
2 \times &({\bf 3}, {\bf 3}, {\bf 1}, 1/3) \cr
         &({\bf 1}, {\bf {\overline 6}}, {\bf 1}, -1/3) \cr
         &({\bf 3}, {\bf 1}, {\bf 2}, 1/6) \cr
         &({\bf {\overline 3}}, {\bf 1}, {\bf 1}, -2/3) \cr
}
\qquad \>\>\>
\eqalign{
& (u \, d)_{\ssc L}, (c \, s)_{\ssc L} \cr
& u^c_{\ssc R}, c^c_{\ssc R} \cr
& d_{3\ssc R}^c   \cr
& (\nu \, l)_{\ssc L} \cr
& l_{\ssc R}^c  \cr
}
\eqalign{
 \> \sim \> &          \cr
 \> \sim \> &          \cr
 \> \sim \> &          \cr
 \> \sim \> &          \cr
 \> \sim \> &          \cr
}
\eqalign{
2 \times &({\bf 1}, {\bf 3}, {\bf 2}, 1/6) \cr
2 \times &({\bf 1}, {\bf {\overline 3}}, {\bf 1}, -2/3) \cr
         &({\bf 1}, {\bf {\overline 3}}, {\bf 1}, 1/3) \cr
3 \times &({\bf 1}, {\bf 1}, {\bf 2}, -1/2) \cr
3 \times &({\bf 1}, {\bf 1}, {\bf 1}, 1) \cr
}
\eqno(assign)
$$
and they of course transform under $\sucppp\times\sucpp\times\suL\times\uY$
in the same way after \(con2) forms.

It is easy to check that all of the gauge anomalies cancel.
This fermion content is designed to break $\sucppp\times\sucpp$ to the
diagonal $\suc$ which is the QCD group of the standard model. This
happens when the quix-antiquix pairs $q_1{\overline q}_1$ and
$q_2 {\overline q}_2$ condense. $\sucppp$ also
produces a top-quark condensate which contributes to electroweak
breaking. (In [\cite{topcolor},\cite{me}], of course, this top-quark condensate
was presumed to be entirely responsible for electroweak breaking. In the
present model, the primary source of electroweak breaking is the
condensate $\langle {\overline\psi}_{\ssc L} \psi_{\ssc R} \rangle$
which enjoys a custodial $SU(2)$ symmetry.)
This gives the top quark a large mass.
We refer the reader to [\cite{me}] for details. This is an example
of how $\xi_{\ss A}$ and $\chi$ can get masses even if the assumptions
leading to \(con3) are false.

\noindent{\it{Variation 4: Breaking of $\tc$
from additional ultracolor interactions.}}
While the idea of $\tc$ breaking itself has a certain economy,
it is also possible to break $\tc$ using a fermion representation
which is vectorial with respect to $\tc$, by introducing some additional
strongly coupled gauge interactions. The simplest version of this
idea is essentially the
same as that used in [\cite{Luty}] and [\cite{Sundrum}]. We introduce
two new strongly coupled groups $SU(n)\times SU(n)^\prime$ (with $n\geq3$)
in addition to $G=\tc \times \sucp \times \suL \times \uYp$. Now
$\tc$ is to be broken by  $SU(n)\times SU(n)^\prime$, but not before
it gets strong enough to produce condensates which break the electroweak
symmetry. The standard model fermions and the technidoublet
$\psi_{\ssc L}, \psi_{\ssc R}^c$ are singlets under
$SU(n)\times SU(n)^\prime$, and transform just as in section 2 under $G$.
This sector of the theory behaves exactly as before. But  instead
of $\eta$, $\xi$, and $\chi$, we introduce fermions transforming
under $SU(n)\times SU(n)^\prime \times \tc \times \sucp \times \uYp$ in
the anomaly-free rep
$$
\eqalign{
& \eta \cr
& \xi  \cr
& \eta^\prime\cr
& \xi^\prime  \cr
}
\eqalign{
\> \sim \> & ({\bf n}, {\bf 1}, {\bf 4}, {\bf 1}, 0)    \cr
\> \sim \> & ({\bf \overline n}, {\bf 1}, {\bf 1},{\bf {\overline 3}},
               { -1/6})
+ ({\bf \overline n}, {\bf 1}, {\bf 1}, {\bf 1}, { 1/2}) \cr
\> \sim \> & ({\bf 1}, {\bf \overline n}, {\bf \overline 4}, {\bf 1}, 0) \cr
\> \sim \> & ({\bf 1}, {\bf n}, {\bf 1}, {\bf 3}, { 1/6})
+ ({\bf 1}, {\bf n}, {\bf 1}, {\bf 1}, { -1/2}) \>\> . \cr
}
\eqno(vtechnis)
$$
Now we assume that $SU(n)$, $SU(n)^\prime$, and $\tc$ all get strong
at roughly comparable scales above the electroweak scale, and that
condensates form according to
$$
\langle \eta^a {\xi}_{\alpha} \rangle = \delta^a_\alpha \, M^3
; \qquad\>\>\>
\langle {\eta}^\prime_{a} \xi^{\prime\alpha} \rangle =
\delta_a^\alpha M^{\prime 3}
\qquad\qquad a,\alpha = 1\ldots 4
\eqno(vcon)
$$
due to $SU(n)$ and $SU(n)^\prime$ respectively. These condensates break
$\tc\times\suPSp \rightarrow \suPS$, so that the
gauged subgroups break according to
$\tc\times\sucp\times\uYp \rightarrow \suc\times\uY$. [As usual we
are using the fact that QCD forces the vacuum to align so as not to break
$\suc$.] Also we may have a condensate
$$
\langle \eta^x \eta_{x^\prime}^\prime \rangle = \delta^x_{x^\prime} m^3
\qquad\qquad x,x^\prime = 1\ldots n
\eqno(conn)
$$
due to $\tc$ which will break
$SU(n)\times SU(n)^\prime$ down to the diagonal $SU(n)$. We are left
in the end with an unbroken $SU(n)$ [or $SU(n)\times SU(n)^\prime$ if
\(conn) does not occur] which confines at energies above
the electroweak scale, and a spontaneously broken $\tc$. The key assumption
here is that the strong coupling dynamics allows
\(con) and \(vcon) to both occur, even though \(vcon) breaks $\tc$.

\noindent{\it{Variation 5: Other technicolor groups.}}
We chose to work with an $SU(4)$ technicolor gauge group
because of the nice way that it can break down to $SU(3)\times U(1)$, thus
serving also as a Pati-Salam group for the techniquarks and technileptons.
It is this structure which allowed us to get one complete standard model
family of technifermions
(after symmetry breaking) from just one technidoublet (before symmetry
breaking). However, it is certainly possible to use other groups. For
example, we could use an $SU(3)_{\ss TC}$, which then breaks with a
weaker $\sucp$ down to the diagonal $\suc$. This group structure was
used in ``Topcolor" models  and in [\cite{Luty}] and
made a cameo appearance in Variation 3 above. The model in
ref.~[\cite{Sundrum}] used an
$SU(5)_{\ss TC}$ which combines with an $\sucp \times U(1)$, leaving
behind the standard model $\suc\times\uY$ as here, but also an unbroken
$SU(2)_{\ss TC}$. This idea of ``hiding" part or all of the technicolor
interaction inside the usual color at low energies can be used also
for larger technicolor groups. In general, one can imagine that part
of the technicolor gauge group is broken completely (e.g.~strong ETC),
part of it combines with some $\sucp$ or $\sucp\times\uYp$ to leave
unbroken the $\suc$ or $\suc \times \uY$ of the standard model, and
part of it remains unbroken and confines some of the technifermions.

\subhead{5. Conclusion}
\taghead{5.}

In this paper, we have discussed some model building ideas
for spontaneously broken technicolor. We should emphasize that any particular
model is subject to assumptions about how and if the condensates form,
since we have no rigorous knowledge about the strong coupling dynamics.
Our experience with QCD is of limited relevance, especially
since we use a chiral representation of the
strongly coupled interaction involving higher dimensional representations.
Can this mutation of the technicolor idea have some beneficial effects?
It has already been argued[\cite{HKOY}] that deconfining technicolor can help
reduce the electroweak $S$ parameter. This will especially be true
if we need only one or several technidoublets. It also seems likely that
breaking technicolor can help to enhance the technifermion condensate scales
compared to the Nambu-Goldstone boson decay constants. This is because
the latter obtain more of a contribution from lower energy scales, where
the masses of the technigluons cut off the technicolor forces. The masses
of the technigluons do not affect the higher energy dynamics which
contribute more to the technifermion condensates. We imagine that a healthy
condensate enhancement compared to the Nambu-Goldstone boson decay constant
could be driven by a combination of a slowly
running technicolor coupling constant and a technigluon mass. Both of
these have the effect of increasing the relative importance of the high
energy dynamics. This has two potentially important effects.
First, the technifermion condensate enhancement might be invoked to suppress
flavor-changing neutral currents.
Second, the technigluons eat Nambu-Goldstone
bosons to get masses which are then proportional to the
Nambu-Goldstone boson decay constants. This could help to explain why the
technigluon masses are small enough to allow condensation in attractive
but subdominant channels involving massless fermions. Perhaps a realistic model
can be constructed using these ideas.

\vskip .2cm

I am grateful to Haim Goldberg, Pierre Sikivie, and Michael Vaughn for
helpful comments. \nugrant

\references

\refis{Slansky}
R.~Slansky, \prpts 79, 1, 1981.

\refis{BHL} W.~A.~Bardeen, C.~T.~Hill, and M.~Lindner,
\journal Phys. Rev., D41, 1647, 1990.

\refis{HLP} C.~T.~Hill, M.~A.~Luty, and E.~A.~Paschos,
\journal Phys. Rev., D43, 3011, 1991.

\refis{Holdom} B.~Holdom,
\journal Phys. Rev., D24, 1441, 1981.

\refis{tumbling} S.~Raby, S.~Dimopolous, and L.~Susskind,
\journal Nucl. Phys., B169, 373, 1980.

\refis{walking} B.~Holdom,
\journal Phys. Lett., 150B, 301, 1985;
V.~A.~Miransky, \journal Nuovo Cimento, 90A, 149, 1985;
T.~Appelquist, D.~Karabali and L.~C.~R.~Wijewardhana,
\journal Phys. Rev. Lett., 57, 957, 1986;
M.~Bando, T.~Morozumi, H.~So and K.~Yamawaki,
\journal Phys. Rev. Lett., 59, 389, 1987.

\refis{ETC} S.~Raby, S.~Dimopolous, and L.~Susskind,
\journal Nucl. Phys., B169, 373, 1980.

\refis{strongETC} T.~Appelquist,  M.~Einhorn, T.~Takeuchi,
L.~C.~R.~Wijewardhana,
\journal Phys. Lett., 220B, 223, 1989;
V.~A.~Miransky, M.~Tanabashi, and K.~Yamawaki,
\journal Phys. Lett., 221B, 177, 1989.

\refis{GMS}
V.~P.~Gusynin, V.~A.~Miransky, and Yu.~A.~Sitenko,
\journal Phys. Lett., 123B, 407, 1983.

\refis{LR} M.~Lindner and D.~Ross,
\journal Nucl. Phys., B370, 30, 1992.

\refis{King} S.~F.~King,
\journal Phys. Rev., D45, 990, 1992.

\refis{KMP} T.~K.~Kuo, U.~Mahanta and G.~T.~Park
\journal Phys. Lett., B248, 119, 1990.

\refis{oblique} D.~C.~Kennedy and B.~W.~Lynn,
\journal Nucl. Phys., B322, 1, 1989;
D.~C.~Kennedy,
\journal Phys. Lett., 268B, 86, 1991.
M.~E.~Peskin and T.~Takeuchi,
\journal Phys. Rev., D46, 381, 1992.

\refis{Luty} M.~A.~Luty, ``Electroweak Symmetry Breaking via a
Technicolor Nambu--Jona-Lasinio model" Lawrence Berkeley Lab preprint
LBL-32089 (1992).

\refis{CGS} R.~S.~Chivukula, M.~Golden, and E.~H.~Simmons,
``Critical Constraints on Chiral Hierarchies", Boston Univ. preprint
BUHEP-92-35 (1992).

\refis{Sundrum} R.~Sundrum, ``A Realistic Technicolor Model from 150 TeV down"
Lawrence Berkeley Lab preprint LBL-32107 (1992).

\refis{HKOY} C.~T.~Hill, D.~C.~Kennedy, T.~Onogi, and H.-L. Yu,
``Spontaneously Broken Technicolor and the Dynamics
of Virtual Vector Technimesons"
Fermilab preprint FERMI-PUB-92/218-T (1992).

\refis{FCNCs} S.~Dimopolous and J.~Ellis,
\journal Nucl. Phys., B182, 505, 1981.

\refis{Nambu} Y.~Nambu,
EFI Report No. 88-39, 1988 (unpublished);
in {\it New Trends in Strong
Coupling Gauge Theories}, 1988 International Workshop,
Nagoya, Japan, edited by M.~Bando, T.~Muta, and K.~Yamawaki),
(World Scientific, Singapore, 1989);
EFI Report No. 89-08, 1989 (unpublished).

\refis{topcolor} C.~T.~Hill,
\journal Phys. Lett., B266, 419, 1991.

\refis{me} S.~P.~Martin,
\journal Phys. Rev., D46, 2197, 1992;
\journal Phys. Rev., D45, 4283, 1992.

\refis{EKR} N.~Evans, S.~King, and D.~Ross,
``Top-Quark Condensation from Broken
Family Symmetry," Southampton preprint, SHEP-91{/}92-11 (1992).

\refis{TC} S.~Weinberg,
\journal Phys. Rev., D19, 1277, 1979;
L.~Susskind, \journal Phys. Rev., D20, 2619, 1979.

\refis{ETC} S.~Dimopolous and L.~Susskind,
\journal Nucl. Phys., B155, 237, 1979;
E.~Eichten and K.~Lane, \journal Phys. Lett., 90B, 125, 1980.

\refis{NJL} Y.~Nambu and G.~Jona-Lasinio,
\journal Phys. Rev., 122, 345, 1961.

\refis{custodial}  P.~Sikivie, L.~Susskind, M.~Voloshin and V.~Zakharov,
\journal Nucl. Phys., B173, 189, 1980.

\refis{alignment}  S.~Weinberg,
\journal Phys. Rev., D13, 974, 1976;
M.~Peskin,
\journal Nucl. Phys., B175, 197, 1980;
J.~Preskill,
\journal Nucl. Phys., B177, 21, 1981.

\endreferences\end